\documentclass[tight,twocolumn]{aastex62}
\usepackage{apjfonts}
\usepackage{psfig}
\usepackage{float}

\usepackage{graphics}
\usepackage{amssymb,amsmath}
\usepackage{color}
\usepackage{todonotes}
\usepackage{xspace}
\usepackage{comment}
\newcommand{\name}{ASASSN-18bt\xspace}

\newcommand{\erg}{\hbox{erg cm$^{-2}$ s$^{-1}$}}

\newcommand{\msun}{\hbox{M$_{\odot}$}}

\newcommand{\halpha}{\hbox{H$\alpha$}}

\newcommand{\sd}{SD\xspace}
\newcommand{\dd}{DD\xspace}
\newcommand{\sn}{SN Ia\xspace}
\newcommand{\sne}{SNe Ia\xspace}

\begin{document}

\title{No Stripped Companion Material in the Nebular Spectrum of the ``Two-Component'' Type Ia Supernova ASASSN-18bt}  
\shorttitle{ASASSN-18bt Nebular Spectrum}
\shortauthors{Tucker, Shappee, and Wisniewski}

\author[0000-0002-2471-8442]{M. A. Tucker}
\altaffiliation{DOE CSGF Fellow}
\affiliation{Institute for Astronomy, University of Hawai'i, 2680 Woodlawn Drive, Honolulu, HI 96822, USA}

\author[0000-0003-4631-1149]{B.~J.~Shappee}
\affiliation{Institute for Astronomy, University of Hawai'i, 2680 Woodlawn Drive, Honolulu, HI 96822, USA}

\author[0000-0001-9209-1808]{J.~P.~Wisniewski}
\affiliation{Homer L. Dodge Department of Physics \& Astronomy, The University of Oklahoma, 440 W. Brooks Street, Norman, OK 73019, USA}

\correspondingauthor{Michael A. Tucker}
\email{tuckerma@hawaii.edu}

\date{Accepted XXX. Received YYY; in original form ZZZ}



\begin{abstract}
We analyze a KeckI/LRIS nebular spectrum taken 268 days after $B$-band maximum of ASASSN-18bt (SN~2018oh), a Type Ia supernova (SN Ia) observed by {\it K2} at the time of explosion. ASASSN-18bt exhibited a two-component rise to peak brightness, possibly the signature of an interaction between the SN ejecta and a large ($\gtrsim 20~R_\odot$) nearby, non-degenerate companion.  We search for emission signatures of stripped material from a non-degenerate companion in the nebular spectrum and find no evidence for any unbound material. We place an upper limit of $< 0.006~M_\odot$ on the amount of stripped/ablated H-rich material that could go undetected in our spectrum, effectively ruling out all hydrogen-rich donor stars. Additionally, we place a more tentative upper limit on HeI emission in the observed spectrum of $\lesssim 0.02~M_\odot$ which also rules out helium star companions. Our deep limits rule out a non-degenerate companion as the explanation for the early-time feature in ASASSN-18bt. 

\end{abstract}


\section{Introduction}

Although Type Ia supernovae (\sne) have been anchors for cosmological studies, their origins remain elusive. The consensus is a carbon/oxygen (C/O) white dwarf (WD) experiencing a thermonuclear explosion \citep{hoyle60}, although the mechanism for actually exploding the WD is still debated. The current literature on \sne progenitors can be grouped into two broad categories: the single degenerate (\sd) and double degenerate (\dd) scenarios (see \citealp{maoz14} for an in-depth review).

The \dd scenario involves two WDs colliding or coalescing to produce a \sn. The exact merger process is unclear, with theories including gravitational wave emission from a close binary \citep[e.g.,][]{tutukov79, iben84, webbink84}, a violent collision due to perturbations from a third \citep[e.g.,][]{thompson11, katz12, shappee13c, antognini14} or fourth body \citep[e.g.,][]{pejcha13},  runaway accretion from the lower-mass WD onto the smaller, more massive WD \citep[e.g.,][]{pakmor12}, or a "Double-detonation", where an initial detonation in an accreted He surface layer triggers carbon detonation in the core of the sub-Chandrasekhar mass WD \citep{fink10, kromer10}. Since both stars involved in this process are intrinsically faint, finding and characterizing these systems is exceptionally difficult \citep[e.g., ][]{rebassa18}. However, some tentative progress has been made, such as bi-modal emission in the nebular phase \citep{dong15}, constraints on nucleosynthetic yields (e.g., \citealp{shappee17}), and potential hyper-velocity remnants \citep{shen18}. No concrete detections of a \dd system have been discovered thus far, and the majority of \dd progenitor conclusions stem from ruling out \sd systems.

In the \sd case, the WD has a non-degenerate companion, such as a main-sequence (MS), sub giant (SG), or red giant (RG) star, undergoing Roche Lobe overflow (RLOF). Mass-transfer onto the WD occurs until the WD is destabilized and explodes.
There are several observational signatures expected from the \sd channel of \sne regardless of the explosion mechanism. X-rays are expected from the accretion onto the WD \citep{lanz05}, which should have observational signatures many years after explosion \citep[e.g., ][]{woods18}. When the ejecta strikes the companion, $\sim 0.15-0.5~M_\odot$ of mass will be stripped/ablated from the donor star \citep{marietta00, pan12, boehner17}, and should be visible once the \sn enters the nebular phase \citep{mattila05, botyanszki18}. Additionally, the ejecta-companion interaction should create a blue signature in the rising light curve detectable within the first $1-2$ days of explosion \citep{kasen10}.  Assuming RLOF, these signatures are dependant on both the radius of the companion and the viewing angle of the explosion, which can mask potential detections. 

Modern surveys have become increasing capable of discovering nearby SNe within a day or two of explosion and probing the early light curves of SNe Ia. There are many \sne which show smoothly-rising light curves  \citep[e.g., ][]{nugent11, cartier17, holmbo18}, yet some exhibit a rise to peak that does not follow a single power law (e.g.,  SN~2012fr \citep{contreras18}, SN~2013dy \citep{zheng13}, SN~2014J \citep{goobar15, siverd15}, MUSSES1604D \citep{jiang17}, iPTF16abc \citep{miller18}, and SN~2017cbv \citep{hosseinzadeh17}) with \name falling into this category \citep{shappee18bt}. This deviation from a single power-law rise has been interpreted as potential ejecta-companion interaction (e.g., SN~2017cbv, \citealp{hosseinzadeh17}), especially when accompanied by a blue excess. Yet searches for other expected signatures of \sd progenitor systems in SN~2017cbv return null results \citep{sand18}. The difference between these two types of early \sne light curves can potentially be explained by a viewing angle argument, but it is still unclear if these deviations are truly indicative of ejecta-companion interaction, or suggestive of other physical processes intrinsic to the \sn explosion. 

\citet{stritzinger18} recently uncovered two distinct \sn populations in observations of \sne within days of explosion, with definitive red and blue populations before converging to a single relation. Furthermore, the presence of a blue excess at early times is correlated with photospheric temperature inferred from spectra at maximum light (``shallow-silicon supernovae''). If the early light curve differences were driven entirely by interaction with a companion and viewing angle, the correlation with \sne photospheric properties at maximum light is puzzling. When reaching maximum light, the ejecta has expanded by a factor of nearly 5000 in volume, is a factor of $\sim 50$ brighter, and is now powered by radioactive decay rather than potentially from shock cooling. Any influence from a possible companion is likely shrouded at this stage in the \sn's evolution. 

Besides early-time light curves, nebular spectroscopy of \sne provide the unique opportunity to place external constraints on the progenitor system. When the companion is struck by the ejecta, mass is liberated from the donor star \citep{wheeler75}. Initially shrouded by the ejecta, this unbound material is obscured until the \sn reaches the nebular phase (i.e., becomes optically thin). 
By placing limits on the non-detection of expected spectral features, statistical limits can be placed on the maximum amount of stripped/ablated material that could go undetected in the observed spectrum. Comparing the derived limits on unbound companion material to dedicated SN-companion interaction simulations in the literature \citep[e.g., ][]{marietta00, pan12, boehner17} can constrain the progenitor system of these \sne. This analysis has been applied to a few dozen \sne in the literature \citep{mattila05, leonard07, lundquist13, shappee13, lundquist15, maguire16, graham17, sand18, shappee18, holmbo18}, with no significant detections of unbound companion material. 

Most previous studies on the amount of unbound material in \sne used the parameterized 1D simulations of \citet{mattila05}, which calculated the expected emission from stripped material at $350~\rm{days}$ after \sn peak brightness. To convert flux/equivalent width upper limits to unbound mass limits, linear scalings were employed \citep[e.g., ][]{leonard07}. Recently, \citet{botyanszki18} utilized multi-dimensional radiative transfer codes to calculate the expected emission from $0.25~M_\odot$ of unbound material. These results differed from those of \citet{mattila05} on two accounts: the amount of clumping in the stripped material, resulting in higher emitted luminsoities, and the scaling between stripped mass and emitted luminosity is closer to exponential than linear. More stringent mass limits can be placed on non-detections of H and He in nebular spectra of \sne with these updated models. Additionally, \citet{botyanszki18} finds that \ion{He}{1} lines should also be present from H-rich unbound material, assuming the stripped material has non-zero metallicity. Thus, \ion{He}{1} emission can also be used to place constraints on both H-rich and He-rich donor stars.


\name (SN~2018oh) was discovered by the All-Sky Automated Survey for SuperNovae (ASAS-SN; \citealp{shappee14}) on 2018 Feb.~4.41 in the {\em K2}  Campaign 16 field \citep{brown18ATel, shappee18bt}.  Since \name was in the K2 field-of-view (FOV) at the time of explosion, the rise is extremely well characterized, especially since this is the brightest \sn observed by Kepler thus far \citep{shappee18bt}, peaking at $B_{\rm{max}} \approx 14.3~\rm{mag}$ on MJD $58162.7\pm0.3$ \citep{li18}. ASASSN-18bt is located in UGC 04780 ($z = 0.010981$, \citealp{schnieder90}) at a mere $52.7\pm1.2~\rm{Mpc}$ \citep{li18}. Additionally, \citet{li18} found no reddening from the host galaxy, so we adopt the Milky Way reddening $E(B-V) = 0.04$~mag \citep{SF11}.  

The mechanism driving the bump and blue excess in the early light curve of \name is under debate. \citet{shappee18bt} compared the rising light curve of \name to various theoretical models including $^{56}$Ni mixing in the ejecta \citep[][]{piro13}, interaction with a RLOF companion \citep{kasen10}, possible interaction with a circumstellar wind \citep[][]{piro16}, and synthetic light curves of double-detonation models \citep{noebauer17}. Based on the shape of the rise, \citet{shappee18bt} concluded that $^{56}$Ni mixing models can span the observed behavior and interaction with a companion is disfavored due to the rapid rise present in models with a large enough signature. However, \citet{dimi18} analyzed a similar set of data and drew a contradictory conclusion, using the observed blue color of \name during the bump as evidence against a double degenerate system. This discrepancy warrants additional constraints on the progenitor system of \name. 

In this Letter, we present a nebular-phase KeckI/LRIS spectrum of \name and place strong limits on the presence of any stripped companion material. In \S\ref{sec:data}, we present our photometric and spectral data acquisition and reduction. In \S\ref{sec:methods}, we outline the our methodology in searching for emission signatures of a SD progenitor system. Finally, in \S\ref{sec:results}, we apply these methods to our nebular spectrum of \name, finding no evidence for stripped/ablated companion material and conclude a likely DD progenitor for \name.

\section{Nebular Spectra and Photometry}\label{sec:data}

\begin{figure}
    \centering
    \includegraphics[width=\linewidth,trim={9.5in 6.5cm 9.5in 6.5cm}, clip]{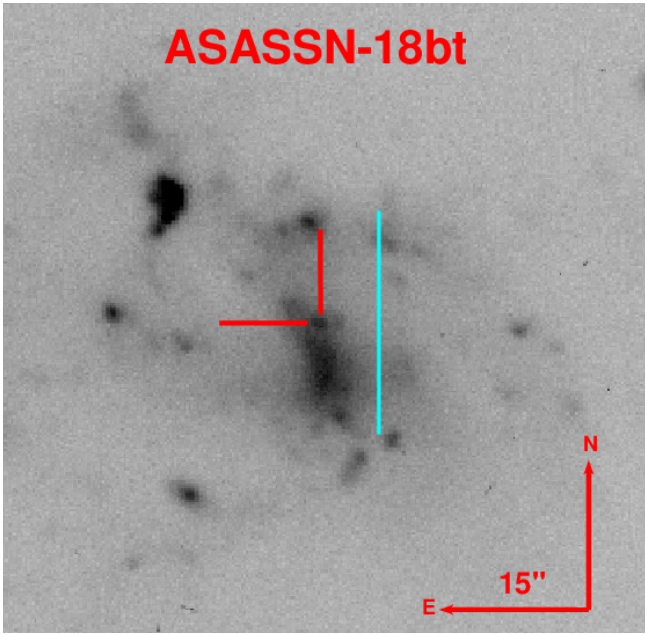}
    \caption{1800~s OSMOS/MDM $r$-band image used for flux-calibrating the nebular spectrum. The location of \name is indicated by the red ticks. The cyan line indicates the length and orientation of the $25^{\prime\prime}~\rm{long}~\times 1^{\prime\prime}$~wide LRIS slit used in the spectroscopic observations. The slit location is shifted horizontally for visual clarity.}
    \label{fig:acqimg}
\end{figure}

We obtained nebular spectroscopy of \name on 2018-11-08 (MJD 58430.65, 268~days after peak $B$-band brightness) with the Low Resolution Imager and Spectrograph \citep[LRIS; ][]{LRISref} on the Keck I telescope using the polarimeter which simultaneously measures the orthogonally polarized components \citep{goodrich95}.\footnote{The polarimeter module in LRIS was in place for other targets observed in the same night, but cannot be removed for individual targets.} Utilizing the $1.\!\!''0$ slit, the reduced spectrum covers roughly $3\,000-10\,000$\AA{} at a nominal resolution of $\sim 7$~\AA{} across the entire spectral range. The individual 2D spectra are reduced with \texttt{Lpipe}\footnote{\url{http://www.astro.caltech.edu/~dperley/programs/lpipe.html}}, implementing typical spectral reduction tasks such as bias subtraction, flat-field correction, wavelength calibration using arc lamp exposures, and flux calibration using spectrophotometric standard stars. The extracted, 1D spectra are finally combined to produce a composite spectrum.  By combining all the individual polarized spectra we create a ``total intensity'' (i.e. unpolarized) spectrum with an effective exposure time of 4800~s, which is shown in the top panel of Fig. \ref{fig:results}. 

Spectrophotometric standard stars are good for relative flux calibration, however, slit losses, weather conditions, and instrumental effects can cause the resulting spectrum to deviate by a factor of a few from absolute flux calibration. Since the analysis in \S\ref{sec:results} depends explicitly on the flux calibration of the observed spectrum we need to place our spectrum on an absolute flux scale.  First we perform aperture photometry on a $1800~\rm s$ $r$-band image acquired on UT 2018-11-03 with the Ohio State MultiObject Spectrograph (OSMOS; \citealp{martini11}) on the MDM Observatory Hiltner 2.4-m telescope and is provided in Fig. \ref{fig:acqimg}. Aperture photometry of ASASSN-18bt is conducted using the IRAF {\it apphot} package and calibrated to the ATLAS \citep{tonry18} All-Sky Stellar Reference Catalog \citep{tonry18b}. 
We measure an $r$-band magnitude of $r = 21.42\pm0.13$~mag, and changing the aperture and sky annulus radii did not substantially affect the resulting magnitude. To place the observed spectrum on an absolute flux scale, we calculate synthetic photometry from the nebular spectrum using Eq. 7 from \citet{fukugita96} then scale the spectrum until the synthetic $r$-band magnitude equals the measured value. We estimate the final flux calibration of the nebular spectrum is good to the precision of the measured $r$-band magnitude ($\sim 15\%$).


\section{The Search for Unbound Companion Material}\label{sec:methods}

\begin{figure*}
\includegraphics[width=\textwidth]{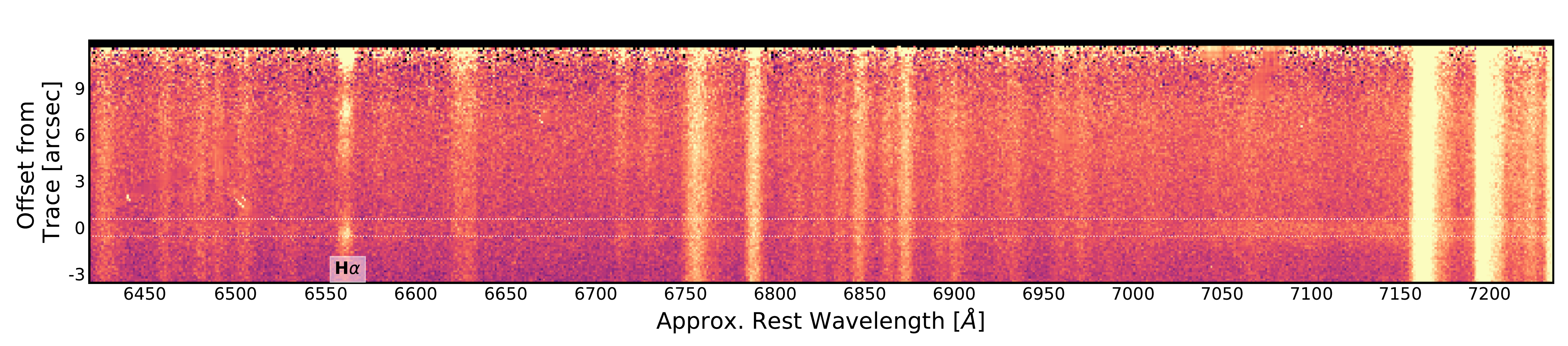}
\caption{Slice of a reduced 2D spectrum of \name. The extraction region of \name is indicated by the faint horizontal white dotted lines. Extended \halpha{} emission is present in the 2D spectrum, indicating the narrow \halpha{} emission in Fig. \ref{fig:results} originates from the host galaxy.
}
\label{fig:2Dspec}
\end{figure*}

\begin{figure}
    \centering
    \includegraphics[width=\linewidth]{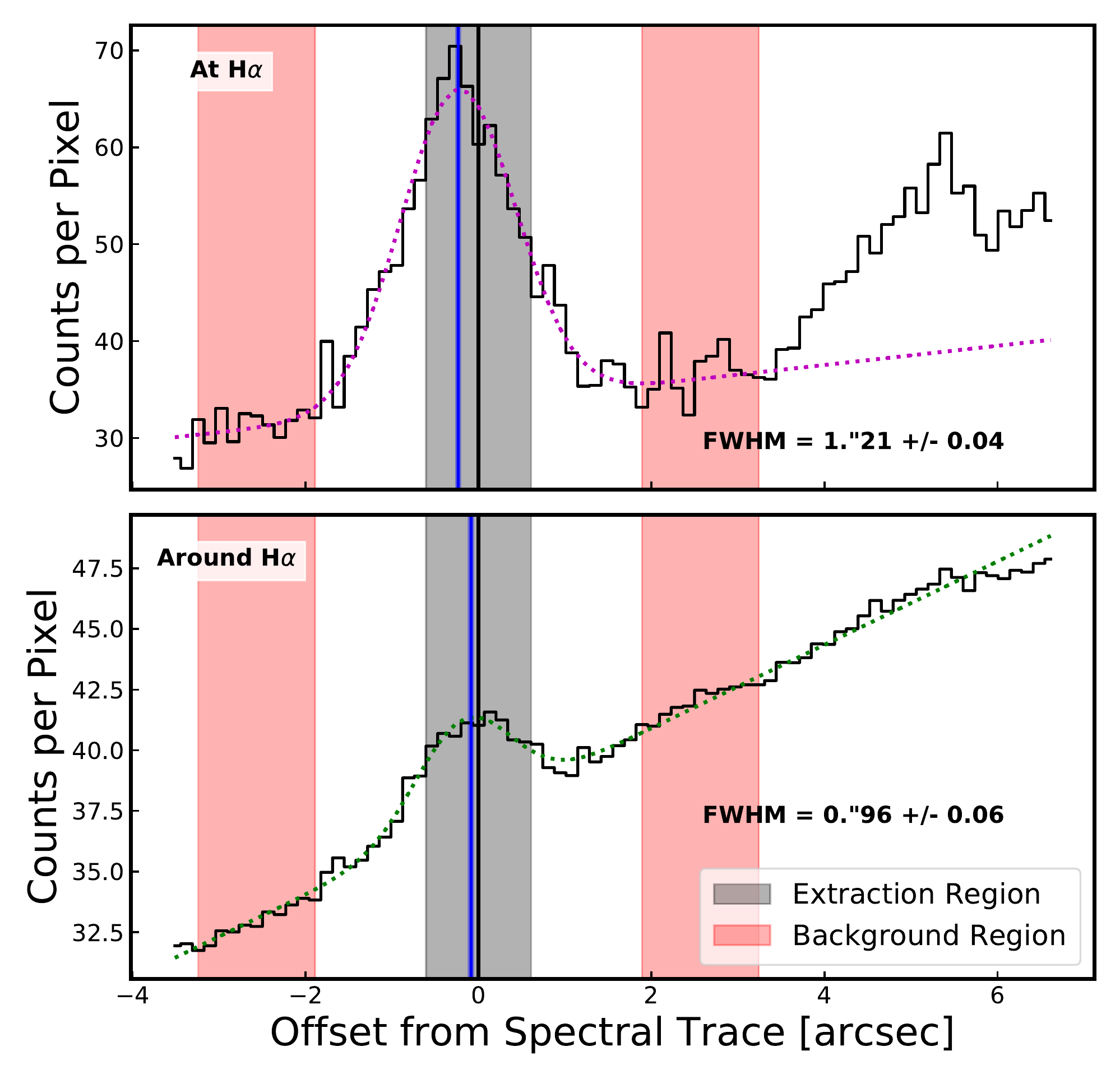}
    \caption{Spatial profiles of \name at \halpha{} (top) and the region surrounding (but excluding) \halpha{} (bottom) in the combined 2D spectra. The vertical black line indicates the center of the spectral trace, whereas blue vertical lines represent the center of the fitted line profile (Gaussian + linear continuum) for each spatial slice. Only pixels within $\pm 3"$ of the spectral trace are included in the line profile fit. The fitted FWHM of each line profile is provided in the bottom right of each panel. The spatial profiles for \name are discrepant at $\sim 3.5\sigma$ in width and $\sim 2.2\sigma$ in line center, indicating the spatial profile at \halpha{} is influenced by host galaxy emission.
    }
    \label{fig:profiles}
\end{figure}

When a \sn explodes in the \sd scenario, the \sn ejecta strikes the companion and removes material from the nearby donor star. 
Hydrodynamic simulations 
agree remarkably well on the amount and velocity distribution of unbound material from the RLOF donor star.  
For MS, SG, RG, and He-star companions we expected $\sim 0.2$, $\sim 0.15$, $\sim 0.5$, and $\sim 0.03$~\msun{} to be removed from the donor star at velocities $\sim750$, $\sim750$, $\sim600$, and $\sim1\,000~\rm{km}~\rm s^{-1}$, respectively. \citep{marietta00, pan12, liu12, liu13, boehner17}

To search for emission from unbound companion material, we follow the methodology of \citet{leonard07} with minor modifications. We fit the continuum with a $2^{\rm{nd}}$-order Savitzky-Golay polynomial \citep{press92} using a smoothing width of $3\,000~\rm{km}~\rm s^{-1}$, which is wider than expected host galaxy lines and narrower than the intrinsic \sn ejecta emission lines. To prevent biasing our derived continuum, we fit the continuum omitting the spectral regions around each expected companion signature. We implement $3\sigma$-clipping in our continuum fitting procedure to prevent instrumental artifacts, host galaxy lines, or telluric features from affecting the measured continuum. 

After fitting the continuum, we subtract it from the observed spectrum and inspect the residuals for emission signatures indicative of material stripped/ablated from a RLOF companion. We inspect the same lines as those predicted by \citet{botyanszki18} for both H-rich and He-rich companions, including the Balmer series (H$\alpha$, H$\beta$, H$\gamma$), and optical \ion{He}{1} lines at 5876\AA{} and 6678\AA{}. We detect no emission lines with the velocity width expected for stripped material ($\sim1\,000~\rm{km}~\rm s^{-1}$).  

We then compute $5\sigma$ upper limits on the equivalent widths (EW) of these lines using Eq. 4 from \citet{leonard01}:

\begin{equation}\label{eq:oldlimit}
    W(5\sigma) = 5\Delta I \sqrt{ W_{\rm{line}}\Delta X}
\end{equation}

\noindent where $W(5\sigma)$ is the $5\sigma$ upper limit on the EW of an undetected spectral feature, $\Delta I$ is the root-mean-square (RMS) around a normalized continuum, $W_{\rm{line}}$ is the width of the spectral feature in \AA{}, and $\Delta X$ is the bin size of the spectrum.
Finally, we translate the EW upper limits into limits on unbound material assuming a distance of $49~\rm{Mpc}$ and using the multidimensional radiative transfer calculations of \citet{botyanszki18} and the decay rate derived by Tucker et al. (in prep).


\begin{figure*}
    \centering
    \includegraphics[width=\linewidth]{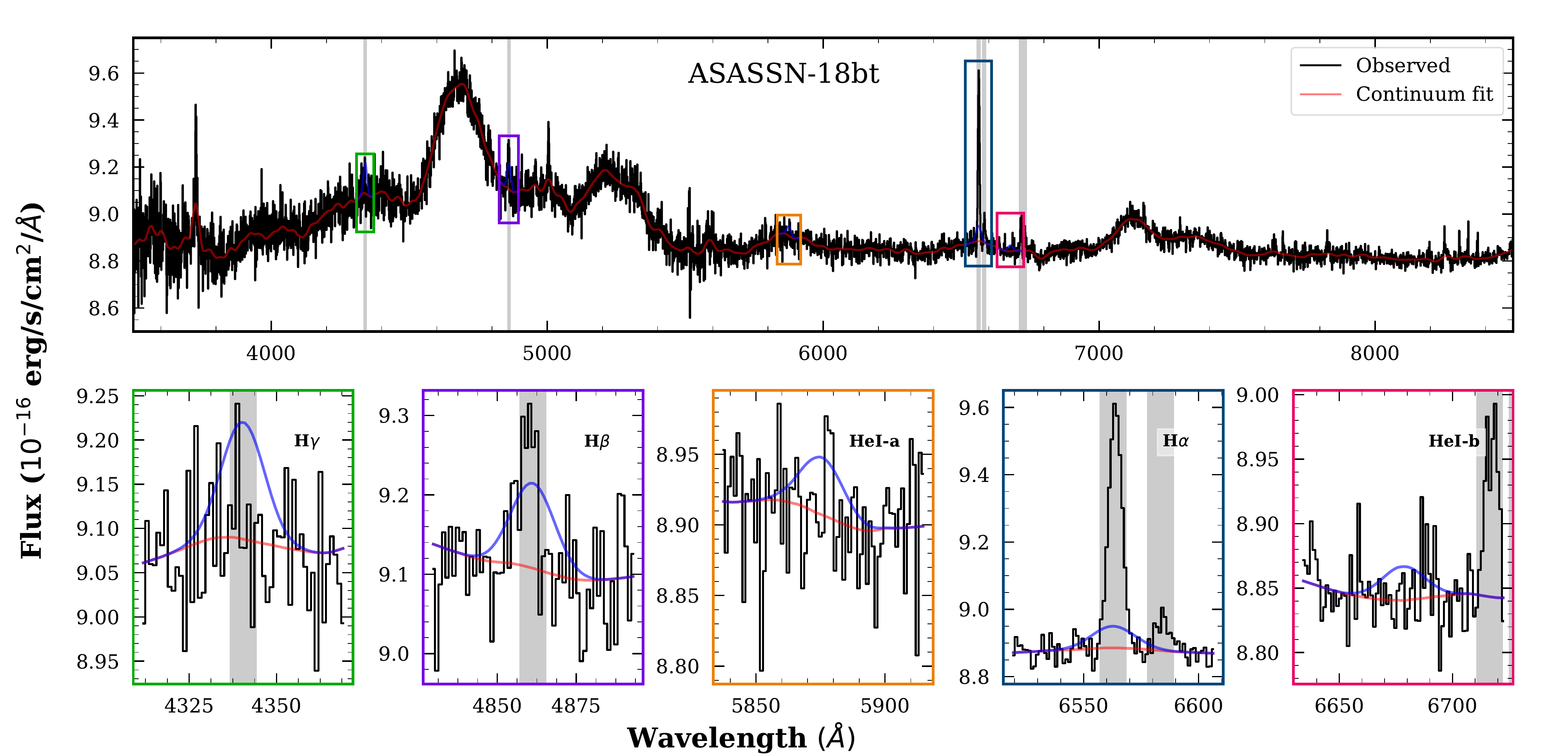}
    \caption{Flux-calibrated nebular spectrum of \name from KeckI/LRIS. \textit{Insets}: Zoomed in view of the region around each expected emission line, denoted in the top right corner, showing the observed spectrum (black), our continuum fit (red), and the empirical flux limit (blue). Axis colors correspond to the boxed locations in the top panel. These flux limits rule out all H-rich RLOF non-degenerate companions and (tentatively) rule out He-rich donors as well (see Table \ref{tab:results}). Gray regions mark masked pixels which are excluded from our continuum fit and the derived $W(5\sigma)$ (see \S\ref{sec:methods}).}
    \label{fig:results}
\end{figure*}

When placing limits on the presence of unbound companion material from the spectrum, underlying host-galaxy emission lines can contaminate the region of interest. Fortuitously, the SN ejecta ($v\sim5\,000 ~\rm{km}~\rm s^{-1}$), stripped/ablated companion material ($v\sim1\,000 ~\rm{km}~\rm s^{-1}$), and host-galaxy emission lines ($v\lesssim100 ~\rm{km}~\rm s^{-1}$) all have significantly different velocities (and therefore line widths). Thus, we are able to disentangle host galaxy and \sn ejecta emission from our derived $W(5\sigma)$, although the instrumental resolution of $\sim 300~\rm{km}~\rm s^{-1}$ places a strict lower limit on any observed velocity profiles. While the host galaxy lines and possible telluric features will be narrower than our continuum smoothing, they will still affect our computed EW limits. 

Host galaxy lines, including the Balmer series, \ion{S}{2}, \ion{O}{3}, \ion{N}{2}, and \ion{Si}{2}, are present in the reduced spectrum of \name. We show a reduced 2D spectrum around H$\alpha$ in Fig. \ref{fig:2Dspec}, which shows apparent narrow, host galaxy \halpha{} emission extended along the spatial axis. While the velocity of this material ($\lesssim 300~\rm{km}~\rm s^{-1}$) is too low to originate from a stripped companion, it may indicate \name stems from a young progenitor system. Further studies of the environment surrounding \name, similar to \citet{lyman18}, are warranted. 

To mitigate contamination from the host galaxy, we mask out regions contaminated by host galaxy lines, telluric effects and instrumental signatures in the reduced spectrum (gray regions in Fig. \ref{fig:results}). 
We implement a correction term to Eq. \ref{eq:oldlimit} to account for these masked regions.  

\begin{equation}\label{eq:correctionfactor}
    f = \sum_{n} G(p_i) / \sum_{i=1}^{N} G(p_i)
\end{equation}

\noindent where $p$ is the set of pixels in the spectral region, $G(p)$ is the Gaussian function computed at pixels $p$, $N$ is the complete set of pixels and $n$ is the \textit{unmasked} subset of pixels used in the limit determination (i.e., $n \subset N$, and $n\equiv N$ when there are no masked pixels in a given spectral region). By definition, $f \in [0,1]~\forall~ p$ and $f < 1$ when any pixels within $2\times \rm{FWHM}$ of line center are masked (i.e., when $n \neq N$).  This accounts for not using all pixels in the spectral region, but weighting each pixel by the Gaussian flux at that pixel since pixels near line center contain more information than pixels on the outskirts of the line. Thus, our $5\sigma$ statistical limit on the non-detection of a given spectral line becomes

\begin{equation}\label{eq:newlimit}
    W(5\sigma) = 5\Delta I f^{-1} \sqrt{ W_{\rm{line}}\Delta X}
\end{equation}

\noindent using the same nomenclature as Eqs. \ref{eq:oldlimit} and \ref{eq:correctionfactor}. This modified form of Eq. \ref{eq:oldlimit} retains the same basic formulation while incorporating the exclusion of contaminated pixels near the expected emission features. Any masked pixels result in $f < 1$, and therefore increase the resulting $W(5\sigma)$, ensuring our statistical limit is robust.

\section{Results}\label{sec:results}

\begin{table*}
    \centering
    \begin{tabular}{c|cccccc}
        Line & $W_{\rm{line}}$ & $W(5\sigma)$ & Flux Limit & Luminosity Limit & H-rich Mass Limit & He-rich Mass Limit \\
         & [\AA] & [\AA] & [$10^{-16}~\erg$] & [$10^{37}~\rm{erg}~\rm s^{-1}$] & [$M_\odot$] & [$M_\odot$]\\\hline
        H$\alpha$ & 21.9 & $0.169$ & $1.50$ & $4.32$ & $0.006$ &  $\ldots$\\
        H$\beta$ & 16.1 & $0.202$ & $1.84$ & $5.29$ & $0.011$ & $\ldots$\\
        H$\gamma$ & 14.5 & $0.225$ & $2.05$ & $5.88$ & $>1$ & $\ldots$\\
        HeI-a & 19.6 & $0.095$ & $0.85$ & $2.44$ & $0.005$ & $0.020$ \\
        HeI-b & 22.3 & $0.071$ & $0.62$ & $1.79$ & $0.007$ & $0.017$ \\\hline
    \end{tabular}
    \caption{Upper limits on undetected spectral features expected from unbound companion material. \ion{He}{1}-a and \ion{He}{1}-b correspond to \ion{He}{1}$\lambda5876$ and \ion{He}{1}$\lambda6678$, respectively.}
    \label{tab:results}
\end{table*}

We compute flux, luminosity, and unbound mass upper limits for each of H$\alpha$, H$\beta$, H$\gamma$, \ion{He}{1}$\lambda5876$, and \ion{He}{1}$\lambda6678$ using Eq. \ref{eq:newlimit}, which are shown in Figure \ref{fig:results} and presented in Table \ref{tab:results}. The gray regions correspond to masked locations in the final spectrum, with prominent masked regions including host galaxy H$\alpha$, H$\beta$, H$\gamma$, \ion{N}{2}, and \ion{S}{2}. 

\subsection{Mitigating Host Galaxy Contamination}

The 1D extracted spectrum of \name includes several regions contaminated by host galaxy emission lines, including \halpha{}. Since \name is a serious \sd system candidate based on the bump and color in the early light curve, all avenues must be explored when considering if the observed \halpha{} emission stems from the host or the \sn itself. We find the probability of the observed \halpha{} emission stemming from a non-degenerate H-rich companion extremely unlikely for the following reasons:

\begin{enumerate}
    \item There is extended \halpha{} emission present in the 2D spectrum (Fig. \ref{fig:2Dspec}).
    \item The host galaxy has observed \halpha{} emission in archival SDSS spectra \citep{SDSS-DR15}.
    \item Narrow \halpha{} emission is also seen in the pre-maximum spectra, and is attributed to the host galaxy \citep{li18}.
    \item The spatial profiles of \name in the 2D spectrum are inconsistent at $2.2\sigma$ in line center and $3.5\sigma$ in line width (Fig. \ref{fig:profiles}).
    \item The derived velocity of the \halpha{} emission line ($\approx 300~\rm{km}~\rm s^{-1}$, the spectral resolution) is several factors too low compared to all dedicated hydrodynamic simulations in the literature \citep{marietta00, pan12, liu12, boehner17}.
    \item There are other host galaxy lines observed in the nebular spectrum, such as \ion{S}{2} and \ion{O}{3}, which have comparable velocity profiles as the observed \halpha{} line and are not expected to be prominent emission lines from unbound material \citep[e.g., ][]{botyanszki18}.
    \item \textit{If} the observed \halpha{} emission was truly from a stripped companion, viewing angle should shift the emission profile shape and center of \halpha{} \citep{botyanszki18} unless the explosion happened while the companion was directly perpendicular to our line of sight.
    \item Any stripped companion material will consist of solar metallicity material, not just pure H. In addition to H emission, this material should exhibit \ion{He}{1} emission \citep{botyanszki18}, for which we have strict constraints (Table \ref{tab:results}).
\end{enumerate}

When masking the host galaxy Balmer emission lines, we first fit the width of the observed H$\alpha$ profile with a measured FWHM of 6.98\AA{} ($\approx 300$ km/s), which is essentially the spectral resolution (i.e., the galaxy lines are unresolved). We mask (in velocity space) the same regions around H$\beta$ and H$\gamma$ to ensure we remove any possible host contamination, even if undetected, in our analysis of the unbound material emission lines. We also apply this method to the host galaxy \ion{N}{2} and \ion{S}{2} emission lines, however, these lines occur near the edges of their respective spectral regions of interest, and are therefore less critical to the final analysis. When computing the flux limit of a spectral region that includes masked portions of the spectrum, we use Eq. \ref{eq:newlimit} as described in \S\ref{sec:methods}.

\subsection{Hydrogen-Rich Companions}
\label{subsec:Hydrogenstar}

The limit on unbound H-rich material derived using H$\alpha$ is slightly less stringent than the same limit derived from the \ion{He}{1}-a line.  However, the modeling for the Balmer series is more extensive than that of the helium emission so we adopt this value for our limit. The strictest $5\sigma$ mass limit on stripped H-rich material, $0.006~M_\odot$ from H$\alpha$, is far lower than the expected amount of unbound mass from simulations in the literature ($\gtrsim 0.15~M_\odot$; \citealp{marietta00, pan12, boehner17}). Our statistical limit concretely rules out the possibility of a H-rich non-degenerate companion, including MS, SG, and RG models. 
Even considering the $3\sigma$ worst case scenario for distance ($58~\rm{Mpc}$) and flux calibration ($r = 21.81~\rm{mag}$), our limits on H-rich unbound material are an order of magnitude stronger than the expected stripped material.    


\subsection{Helium Star Companions}
\label{subsec:Hestar}

In addition to H-rich non-degenerate companions, helium star donors have also been proposed as possible progenitor systems for \sne. Helium stars are typically the result of binary interaction, making them reasonable candidates for \sn progenitors. In the literature, there have been a few studies on possible effects of SN ejecta impacting a nearby RLOF He star. \citet{pan12} and \citet{liu13} performed hydrodynamic simulations modeling the interaction between the SN ejecta and the stellar surface, finding $0.023-0.057~M_\odot$ of material is removed from the donor star by the explosion (He-r model from \citealp{pan12} and W7\_He01/2\_r models from \citealp{liu13}). 

\citet{botyanszki18} also performed non-LTE spectral modeling to predict line luminosities from He-rich unbound material. However, \citet{botyanszki18} does not self-consistently implement the velocity and density distributions from helium star hydrodynamic simulations, but instead replaces the unbound mass from the MS model with pure helium. We inherently assume the same scaling between mass and luminosity for He-rich material as H-rich material, which is reasonable and likely more accurate than 1-to-1 scaling relations previously assumed in the literature. However, we conservatively view our  He-rich mass limits as tentative, until more detailed theoretical studies are conducted. 

Our results place strong statistical limits on the amount of He-rich unbound companion material, with $5\sigma$ upper limits of $0.020~M_\odot$ and $0.017~M_\odot$ for \ion{He}{1}$\lambda5876$ and \ion{He}{1}$\lambda6678$, respectively. These mass limits rule out even the least stripped helium star models discussed above.
Furthermore, even in the absence of our limit, we note that a helium star companion would be too small ($\lesssim 0.4~R_\odot$, see \citealp{liu13}) to explain the large early-time blue excess observed in the K2 light curve of \name. 

\section{Conclusion and Discussion}

In this Letter, we present the first nebular spectral analysis of \name, a \sne observed by \textit{K2} moments after explosion. Although pre-peak data for \name is exquisite in terms of precision and temporal sampling, the interpretation is inconclusive. One of the most enigmatic features of \name is the two-component rising light curve \citep{shappee18bt, dimi18}.  Previously, similar features have been interpreted as potential ejecta-companion interaction \citep[e.g., ][]{hosseinzadeh17}. The \citet{shappee18bt} analysis of the rising \textit{K2} light curve of \name favors shallow $^{56}$Ni models over companion interaction models, whereas the analysis conducted by \citet{dimi18} concludes a likely \sd progenitor system as the origin of the bump and blue excess. This discrepancy warrants additional independent constraints on \name's progenitor system. 

With our nebular-phase KeckI/LRIS spectrum acquired $268$ days after $B$-band maximum, we constrain the amount of H and He emission from possible unbound companion material. Using available models in the literature \citep{botyanszki18}, we convert our non-detection of H and He emission lines into upper limits on the amount of unbound mass. The resulting H and He mass limits rule out non-degenerate H-rich and (tentatively) He star companions (Fig. \ref{fig:results}, Table \ref{tab:results}).  

Since a \sd progenitor system does not match the observed early- and nebular-phase constraints, we consider a \dd system as the likely progenitor of \name. However, there is still the question of what causes the initial rise in the early light curve.  As discussed in \citet{shappee18bt}, a small amount of $^{56}$Ni in the outer layers of the ejecta may reproduce the observed early light-curve component.
Furthermore, there is possible direct evidence for shallow $^{56}$Ni in the nearby SN~2014J; a SN Ia which also exhibited an early rise which cannot be explained by a single power law \citep{goobar15, siverd15}. \citet{diehl14} and \citet{isern16} claim detections of the 158 keV $^{56}$Ni gamma-ray decay lines in SN~2014J between 16-35 days after explosion, requiring $\sim 0.05$ \msun{} of shallow $^{56}$Ni.

Early-time SNe Ia light curves have revealed unexpected diversity, giving us additional views into their progenitor systems.  Yet, while rapid progress has been made, there are still large uncertainties in our understanding of these progenitors systems.  This uncertainty highlights the continuing need for increased theoretical and observational work.

\vspace{1cm}

\textit{During the review process for this manuscript, \citet{dimineb} released a similar, independent paper on nebular spectra of \name. The findings of both studies are in agreement, each placing strict upper limits on the possibility of any unbound companion material.}

\vspace{1cm}
\section*{Acknowledgements}

We thank the referee for their useful comments and improvement of the manuscript. We thank Justin Rupert (MDM), Joshua Shields, and Krzysztof Stanek (Ohio State University) for acquiring the finding MDM $r$-band image. We thank Gagandeep Anand, Connor Auge, Aaron Do, Ryan Foley, Anna Payne, Jose Prieto, and Jennifer van Saders for useful discussions.  MAT acknowledges support from the United States Department of Energy through the Computational Sciences Graduate Fellowship (DOE CSGF).

\bibliographystyle{mnras}
\bibliography{references}
\end{document}